\documentclass[epj]{svjour}
\usepackage{epsfig}
\usepackage{amssymb}
\usepackage{amsmath}
\allowdisplaybreaks[2]

\renewcommand{\vec}[1]{{\mathbf{#1}}}
\newcommand{\Zs}{Z${}_\sigma$}
\newcommand{\Es}{E${}_\sigma$}
\newcommand{\DO}{$D1$}
\newcommand{\DOs}{$D1S$}
\newcommand{\Ecm}{\mbox{$E_{\rm c.m.}$}}
\newcommand{\Ecmdir}{\mbox{$E_{\rm c.m.}^{\rm dir}$}}
\newcommand{\Ecmfit}{\mbox{$E_{\rm c.m.}^{\rm fit}$}}
\newcommand{\Ecmosc}{\mbox{$E_{\rm c.m.}^{\rm osc}$}}
\newcommand{\Ecmmic}{\mbox{$E_{\rm c.m.}^{\rm mic}$}}
\newcommand{\Etot}{\mbox{$E_{\rm tot}$}}
\newcommand{\Psqr}{\mbox{$\langle \hat{\vec{P}}^2_{\rm c.m.} \rangle$}}
\newcommand{\cmc}{c.m.\ correction}
\newcommand{\asurf}{\mbox{$a_{\rm surf}$}}
\newcommand{\avol}{\mbox{$a_{\rm vol}$}}
\newcommand{\asym}{\mbox{$a_{\rm sym}$}}
\newcommand{\rmd}{{\rm d}}
\newcommand{\iunit}{{\rm i}}
\newcommand{\nn}{\nonumber}
\newcommand{\gk}{\gtrless}  
\newcommand{\Kappa}{K}
\newcommand{\MeV}{\mbox{MeV}}

\begin{document}

\title{Consequences of the center--of--mass correction\\
       in nuclear mean--field models}

\author{M. Bender\inst{1,2,3}     \and
        K. Rutz\inst{3}           \and
        P.--G. Reinhard\inst{4,5} \and 
        J. A. Maruhn\inst{3,5}
}

\institute{Department of Physics and Astronomy,
           The University of Tennessee,
           Knoxville, TN 37996, U.S.A.
           \and
           Department of Physics and Astronomy,
           The University of North Carolina,
           Chapel Hill, NC 27516, U.S.A.
           \and
           Institut f\"ur Theoretische Physik, 
           Universit\"at Frankfurt,
           Robert--Mayer--Str. 10, 
           D--60325 Frankfurt am Main, Germany
           \and
           Institut f\"ur Theoretische Physik II, 
           Universit\"at Erlangen--N\"urnberg,
           Staudtstr. 7, 
           D--91058 Erlangen, Germany
           \and
           Joint Institute for Heavy--Ion Research, 
           Oak Ridge National Laboratory,
           P.O. Box 2008, Oak Ridge, TN 37831, U.S.A.
}

\date{October 8, 1999}

\abstract{
We study the influence of the scheme for the correction for spurious
center--of--mass motion on the fit of effective interactions 
for self--consistent nuclear mean--field calculations.
We find that interactions with very simple center-of-mass
correction have significantly larger surface coefficients than 
interactions for which the center--of--mass correction was calculated for 
the actual many--body state during the fit. The reason for that is
that the effective interaction has to counteract the wrong trends with
nucleon number of all simplified schemes for center--of--mass correction
which puts a wrong trend with mass number into the effective interaction 
itself. The effect becomes clearly visible when looking at the deformation 
energy of largely deformed systems, e.g.\ superdeformed states or fission 
barriers of heavy nuclei.
}


\maketitle
%
%
\section{Introduction}
It is generally known that the ground--state wave functions
of mean--field models break symmetries which had been originally given 
in the many--body Hamiltonian or effective energy functional.
Violation of translational symmetry is unavoidable because
the center--of--mass of a system is localised by the mean--field potential. 
This causes a spurious contribution from the center--of--mass
vibrations to the energy and other observables. The problem has been
discussed since decades and several solutions have been developed in
the course of time, for an overview see \cite{Ringbook}. A rigorous 
way to restore the broken symmetries is the projection method.
Projection--before--variation is the perfect solution which has been
applied even in realistic applications \cite{MONSTER}, but it
constitutes a numerically extremely challenging task and is still too
costly to be used in large--scale investigations of nuclear
structure. A simpler approach is the projection--after--variation
method where the mere HFB state is varied but projected wave functions
are used to calculate observables \cite{Ringbook}. A study of this
approach in the context of self--consistent models has hinted that full
projection effects could be quantitatively important in light nuclei
\cite{Schmid}. Nonetheless, by far the most applications deal with
approximate ways to compute the center--of--mass corrections for reasons
of feasibility and transferability. The standard procedure is to
expand the correction in orders of moments 
$\langle\hat{\bf P}{}_{\rm c.m.}^{2n}\rangle$ and to stop at first order
\cite{Blaizot}. And even that is often further simplified in various
manners. As a consequence there are several recipes around for
performing the center--of--mass correction.

This diversity as such could possibly be bearable. The situation is
complicated by the fact that all quantitatively successful nuclear
mean--field theories employ a phenomenological adjustment of the model
parameters \cite{SkyrmeFit}.
While fitting the parameters one has to decide for one of the current
forms of the center--of--mass correction. The first rule to be obeyed is
then that all later applications should employ precisely that recipe
which had been used during the fit \cite{SkyrmeFit}, but the
influence of the actual recipe goes further than that. The various
approximations in themselves do have slightly different trends with
mass number. These differences can be counterweighted to a certain
extend by slight readjustments within the model parameters relevant
for those trends, but the flexibility of the models is limited and
systematic differences appear for larger extrapolations, in case
of the center--of--mass correction the
computation of nuclear--matter properties and finite nuclei at
large deformation, e.g.\ in superdeformed states or fission. 
It is the aim of this paper to present
a thorough investigation of these subtle side effects from the different
recipes for the correction for spurious center--of--mass motion
which will be simply denoted  as \cmc\ in the following. There are,
however, several other corrections for spurious motion and broken
symmetries to be made, but the \cmc\ is among the most important 
ones since it is present in all nuclei, while the influence of, 
e.g.\ the rotational and 
vibrational corrections can be suppressed by choosing spherical nuclei
with stiff potential energy surfaces. In those, the rotational correction
vanishes and the admixture of vibrational excitations to the 
independent-quasiparticle ground state can be assumed 
to be negligible.

The paper is outlined as follows: In Section \ref{Sct:Frame} we
explain briefly the underlying mean--field models in this study. In
Section \ref{sec:cmc} we summarise the currently used approximations
for the center-of-mass corrections and in Section \ref{sec:result},
we present and discuss typical observables we find to be sensitive
to the treatment of the correction for center-of-mass motion:
binding energy systematics and deformation energies. An
Appendix presents the formulae needed to calculate the \cmc\
in relativistic and non--relativistic models.
%
%
\section{Framework}
\label{Sct:Frame}
We investigate the \cmc\ in the frameworks of the self--consistent 
Skyrme--Hartree--Fock (SHF) approach \cite{SHFrev} and the 
relativistic mean--field model (RMF) \cite{Rei89a,Rin95a}. 
In both, SHF and RMF, models the corrections for spurious motion 
can be treated non--relativistically.

The numerical procedure represents the coupled SHF and RMF equations 
on a grid in coordinate space using a Fourier definition of the 
derivatives and solves them with the damped gradient iteration 
method \cite{dampgrad}. We consider both spherical and axially 
symmetric deformed configurations.

Pairing correlations are treated in the BCS scheme using a delta pairing 
force \cite{Krieger} $V_{\rm pair} = V_q \, \delta({\bf r}_1 - {\bf r}_2)$. 
The pairing strengths $V_{\rm p}$ for protons and $V_{\rm n}$ for 
neutrons depend on the actual mean--field parametrisation. They are
optimised by fitting for each parametrisation separately the pairing gaps 
from a fourth--order finite--difference formula of binding energies
in isotopic and isotonic chains of semi--magic nuclei throughout the
chart of nuclei. The pairing--active space is chosen to include the 
number of one additional shell of oscillator states above the Fermi energy
with a smooth Fermi cutoff weight, for details see \cite{pairStrength}.
%
%
\section{The center--of--mass correction}
\label{sec:cmc}
The \cmc\ -- the change in binding energy from projection--after--variation 
in first--order approximation -- is given by
\begin{equation}
\label{eq:Ecm:mic}
\Ecmmic
= - \frac{1}{2mA} \Psqr
\quad .
\end{equation}
We will denote this as \emph{microscopic \cmc} throughout this paper,
referring to the fact that it is calculated from the actual many--body
state. The explicit dependence on mass number of (\ref{eq:Ecm:mic}) 
might cause some trouble when calculating mass differences, see
\cite{Dob95a}. In the framework of energy density functionals, 
the factor $A$ should be replaced by the integral
of the local isoscalar density; since we will use
(\ref{eq:Ecm:mic}) for correction--after--variation only 
throughout this paper, however, this would make no difference.

\mbox{$\hat{\bf P}_{\rm cm} = \sum_k \hat{\bf p}_k$} 
is the total momentum operator in the center--of--mass frame, which
is given by the sum of the single--particle momentum operators. 
Although the BCS state has vanishing
total momentum \mbox{$\langle \hat{\bf P}_{\rm cm} \rangle = 0$}, 
it is not an eigenstate of $\hat{\bf P}_{\rm cm}$, but
has a non--vanishing expectation value of its square
\begin{eqnarray}
\label{eq:BCSPsqr}
\Psqr
& = &   \sum_{\alpha} v_\alpha^2 \; p^2_{\alpha \alpha}
      - \sum_{\alpha, \beta} v_\alpha^2 v_\beta^2 \; 
        {\bf p}_{\alpha \beta} \cdot {\bf p}^{*}_{\alpha \beta}
      \nonumber \\
&   & \quad
      + \sum_{\alpha, \beta} v_\alpha u_\alpha v_\beta u_\beta \; 
        {\bf p}_{\alpha \beta} \cdot {\bf p}_{\bar{\alpha} \overline{\beta}} 
\quad .
\end{eqnarray}
The $\alpha$ and $\beta$ denote single-particle states.
The $p^2_{\alpha \alpha}$ are single--particle expectation values of the 
square of the single--particle momentum operator. They appear only in 
the direct term of the correction. The ${\bf p}_{\alpha \beta}$ 
are off-diagonal single--particle matrix elements of the 
momentum operator. Their squares result from the exchange terms in
$\langle \hat{\bf P}{}^2_{\rm cm} \rangle$. The further evaluation
of (\ref{eq:BCSPsqr}) is outlined in the Appendix.

Although \Ecmmic\ as given by (\ref{eq:Ecm:mic}) 
is already the first--order approximation for the 
momentum--projected binding energy there exist
numerous further approximations for \Ecmmic\ in the literature:
\begin{description}
\item[(A)] 
      The full correction (\ref{eq:BCSPsqr}) is considered in 
      the variational equation and the calculation of the 
      binding energy $\Etot$
      \begin{equation}
      \delta ( {\cal E}_{\rm int} - \Ecmmic ) 
      = 0
      \quad , \quad
      \Etot
      = {\cal E}_{\rm int} - \Ecmmic
      \quad .
      \end{equation}
      This was employed so far only in the fit of the Skyrme interactions
      SLy6 and SLy7 \cite{Chabanat,SLyx}. For HF states without pairing
      \Ecm\ gives an additional term to the equa\-ti\-ons--of--motion
      but the HF equations can be solved as usual. For HFB states the 
      mean field becomes state dependent, which requires 
      an additional constraint on orthonormal single--particle wave functions 
      in the variational equation as described in \cite{ourHFB}. 
      The numerical solution of the resulting equations of motion is very 
      costly, especially in deformed calculations. Therefore this scheme 
      was employed so far only for the description of doubly--magic nuclei 
      where an HF state can be used.
\item[(B)]
      The \cmc\ is omitted in the variational equations, but the 
      microscopic correction (\ref{eq:BCSPsqr}) is considered when 
      calculating the total binding energy
      \begin{equation}
      \delta {\cal E}_{\rm int} 
      = 0
      \quad , \quad
      \Etot
      = {\cal E}_{\rm int} - \Ecmmic
      \quad .
      \end{equation}
      This \emph{a posteriori} correction scheme is used, e.g., for the 
      Skyrme forces SkI1--SkI5 \cite{SkIx} and the RMF forces 
      NL--Z \cite{NLZ}, and PL--40 \cite{PL40}.
\item[(C)] 
      The \cmc\ is approximated by its diagonal (direct) terms 
      \begin{equation}
      \Ecmdir =   \frac{1}{2mA} \sum_{k} v_k^2 \; p^2_{kk}
              = - \frac{\hbar^2}{2mA} \int \! \rmd r \; \tau
      \end{equation}  
      -- where $\tau$ is the local kinetic density -- 
      but employed before variation
      \begin{equation}
      \label{eq:cmcorr:diag}
      \delta ( {\cal E}_{\rm int} - \Ecmdir ) 
      = 0
      \quad , \quad
      \Etot
      = {\cal E}_{\rm int} - \Ecmdir
      \quad .
      \end{equation}
      The contribution from the \cmc\ to the (non--relativistic) 
      equations of motion has the same structure as the kinetic term 
      and leads to a renormalisation of the mass of the nucleons
      \begin{equation}
      \frac{1}{m}
      \rightarrow \frac{1}{m} \left( 1 - \frac{1}{A} \right)
      \quad .
      \end{equation}
      This scheme is used for most Skyrme interactions like 
      SIII \cite{SIII}, SkM* \cite{SkM*}, SkP \cite{SkP}, 
      the Skyrme forces of Tondeur \cite{Tx}, SLy4 and SLy5 \cite{SLyx},
      the Skyrme interactions used in semi--microscopic ETFSI 
      calculations \cite{ETFSI}, and the Gogny forces \cite{Gogny}.

%
%
\begin{table}[t]   
\caption{\label{NucMat}
Compilation of nuclear matter properties for a number of typical 
parameter sets. SkM*--SLy6 are Skyrme forces, \DO\ and \DOs\ are Gogny 
interactions, and NL1, NL--Z, NL3, and NL--SH RMF forces. 'scheme' is 
the scheme for \cmc\ employed in the fit of the particular interaction, 
\protect\avol\ denotes the volume coefficient or energy per nucleon, 
\protect\asym\ the (volume) symmetry coefficient, 
and \protect\asurf\ the surface coefficient. Empirical values for the
volume coefficients derived from the liquid--drop model are 
\mbox{$\avol = -16.0 \pm 0.2$} and \mbox{$\asym = 32.5 \pm 0.5$}
\protect\cite{Cha97a}.
}
\begin{center}
\begin{tabular}{lccccc}
\hline\noalign{\smallskip}
Force & Ref.
      & scheme
      & \protect\avol
      & \protect\asym   
      & \protect\asurf \\
      &
      &
      & $[{\rm MeV}]$
      & $[{\rm MeV}]$ 
      & $[{\rm MeV}]$  \\
\noalign{\smallskip}\hline\noalign{\smallskip}
SkM*  & \cite{SkM*}   & C & $-15.75$ & 30.0 & 17.6 \\
SkP   & \cite{SkP}    & C & $-15.92$ & 30.0 & 18.0 \\ 
SkT6  & \cite{Tx}     & C & $-15.94$ & 30.0 & 18.1 \\
\Es   & \cite{SkyrmeFit} & C & $-16.00$ & 26.5 & 18.2 \\ 
\Zs   & \cite{SkyrmeFit} & B & $-15.85$ & 26.7 & 16.9 \\
SkI1  & \cite{SkIx}   & B & $-15.93$ & 37.5 & 17.3 \\ 
SkI3  & \cite{SkIx}   & B & $-15.96$ & 34.8 & 17.5 \\ 
SkI4  & \cite{SkIx}   & B & $-15.92$ & 29.5 & 17.3 \\
SLy4  & \cite{SLyx}   & C & $-15.97$ & 32.0 & 18.2 \\
SLy6  & \cite{SLyx}   & A & $-15.92$ & 32.0 & 17.4 \\ 
\noalign{\smallskip}\hline\noalign{\smallskip}
\DO   & \cite{Gogny}  & C & $-16.02$ &      & 20.3 \\
\DOs  & \cite{Ber89a} & C & $-16.02$ &      & 18.2 \\
\noalign{\smallskip}\hline\noalign{\smallskip}
NL1   & \cite{NL1}    & E & $-16.42$ & 43.5 & 18.7 \\
NL-Z  & \cite{NLZ}    & B & $-16.19$ & 41.7 & 17.7 \\ 
NL3   & \cite{NL3}    & D & $-16.24$ & 37.4 & 18.5 \\
NL-SH & \cite{NLSH}   & D & $-16.35$ & 36.1 & 19.1 \\
\noalign{\smallskip}\hline\noalign{\smallskip}
\end{tabular}
\end{center}
\end{table}
%
%
\item[(D)] 
      The microscopic \cmc\
      (\ref{eq:Ecm:mic}) can be evaluated analytically for
      harmonic oscillator states. Using the usual parameterisation of the
      oscillator constant from the Nilsson model one obtains an estimate
      for (\ref{eq:Ecm:mic}) as 
      \begin{equation}
      \label{eq:Ecm:HOest}
      \Ecmosc
      = - \tfrac{3}{4} \, 41 \; A^{-1/3} \; {\rm MeV}
      \quad .
      \end{equation}
      This is used for the RMF forces NL--SH \cite{NLSH}, TM1 \cite{TM1}, 
      and NL3 \cite{NL3}. Because
      (\ref{eq:Ecm:HOest}) does not depend on the many--body wave function
      this gives no contribution to the variational equations.
\item[(E)]
      The \cmc\ is approximated by
      \begin{equation}
      \label{eq:Ecm:fit}
      \Ecmfit
      \approx - 17.2 \; A^{-0.2} \; {\rm MeV}  
      \end{equation}
      which is a fit to values of the full \emph{a posteriori} 
      correction for the 
      fit nuclei calculated with the Skyrme interaction \Zs\
      \cite{SkyrmeFit}. This is used for the RMF forces NL1 
      and NL2 \cite{NL1}.
\end{description}
Some of these approximate schemes for \cmc\ are discussed and compared
in \cite{But84a}, where also an additional one is proposed which is
not used in the models discussed here and a short historical overview 
is given. However, the authors of \cite{But84a} restrict themselves to
the quality of the various schemes to approximate (\ref{eq:Ecm:mic}).

As can be seen from the harmonic oscillator value (\ref{eq:Ecm:HOest}) 
-- which will turn out to give at least the right order of magnitude for 
\Ecmmic\ in heavy nuclei -- the \cmc\ decreases with increasing mass 
number $A$ and vanishes for infinite homogeneous nuclear matter 
(which is translationally invariant). Since the total binding energy 
increases with $A$, the relative contribution of the \cmc\ to the 
total binding energy is largest for very small nuclei.

In the correction--before--variation schemes (A) and (C) the \cmc\ 
directly affects other observables than the total binding 
energy as well, for example single--particle energies. Like 
in case of the total binding energy the effect of the correction 
decreases with increasing $A$. All other schemes deliver an
approximate correction of the binding energy only, but analogous 
a posteriori correction of order $\langle\hat{\bf P}{}_{\rm c.m.}\rangle$ 
can be performed for the local density $\rho (\vec{r})$ and with 
that all observables derived from it \cite{Rei89a,ReiCode}.

We mention in passing that no \cmc\ of the binding energies is performed
in macroscopic--microscopic models, but the smooth part of \Ecm\ is 
implicitly contained in the macroscopic part of these models.
%
%
\section{Results and Discussion}
\label{sec:result}
%
%
\subsection{Nuclear Matter Properties}
\label{subsect:res:nucmat}
From the model systems of infinite nuclear matter (INM) and 
semi-infinite nuclear matter (SINM) at saturation density one 
obtains the leading terms in the nuclear liquid-drop mass formula
\begin{equation}
\label{eq:LDM}
E
=   \avol \, A
  + \asym \, I^2 \, A  
  + \asurf \, A^{2/3}
  + \ldots
\quad ,
\end{equation}
where $A$ is the mass number and $I$ the relative neutron excess 
given by \mbox{$I = (N-Z)/A$}. The relation between 
self--consistent mean--field models and the liquid--drop model and 
its refinements like the droplet model
is discussed in \cite{Brack85}. While the 
volume coefficient \avol\ and volume symmetry coefficient \asym\
calculated in INM are directly comparable with the liquid--drop
values, the surface coefficient \asurf\ has to be handled more 
carefully. Usually \asurf\ is extracted from 
calculations of SINM in semi--classical approximation
which leads to a value of the surface coefficient
that is not directly comparable to the liquid--drop model value
\cite{Brack85,Far81a,Pea82a}. Instead the particular Skyrme 
interaction SkM* -- which is fitted with emphasis on its surface
properties -- is often used as a reference when discussing
the surface coefficient of effective interactions.

The INM properties of effective interactions have been extensively 
discussed elsewhere, see e.g.\ 
\cite{Rei89a,SHshells,Brack85,Bla95a,Cha97a}
and references therein. The properties of SINM have gained a
lot of attention as well, see e.g.\ 
\cite{SkM*,Tx,Brack85,Far81a,Pea82a,Far82a,Far84a,Far97a}
for non--relativistic models and
\cite{Cen93a,Cen93b,DEs97a,Spe93a,VEi94b,VEi94a,Sto97a,DEs99a}
for relativistic interactions.
We just want to repeat the findings relevant for our study,
Table~\ref{NucMat} summarises the coefficients of (\ref{eq:LDM}) 
as calculated from symmetric INM and SINM for a number of effective 
interactions. Most non--relativistic (relativistic) 
interactions agree among each other in the values for \avol\ 
and \asym\ although there are some minor differences and a 
few exceptions, but relativistic and non--relativistic 
interactions differ significantly, the RMF forces give \avol\
and \asym\ larger than the empirical values, while the Skyrme
forces work much better in that respect. It is important to note,
however, that not all the values for Skyrme forces 
are predictions. For some parametrisations, i.e.\ SLy4, SLy6, 
SkP, SkT6 and SkM*, INM properties were used as input data 
during their fit.
%
%
\begin{figure}[t!]      
\epsfig{file=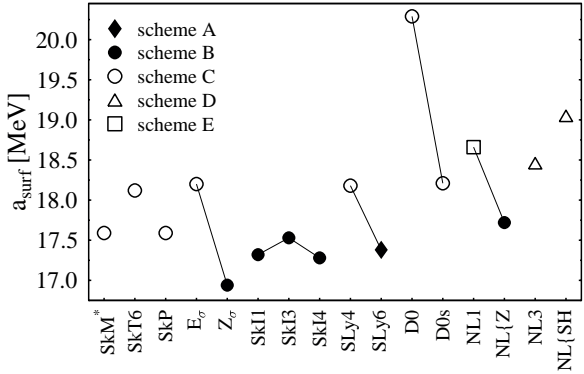}
\caption{\label{asurf} 
Surface coefficient \asurf\ for a variety of effective
interactions. Lines connect interactions which are similarly
fitted. Open markers denote forces with a simple scheme for 
\protect\cmc, while filled markers denote interactions where
\Ecm\ is calculated microscopically within scheme (A) or (B).
The interactions with microscopically calculated \Ecm\ have 
significantly smaller surface coefficients.
}
\end{figure}
%
%

The values for the surface coefficient \asurf\ differ
significantly among all effective interactions,
see also Figure~\ref{asurf}. We present here values calculated
with the extended Thomas--Fermi method taken from \cite{Sto97a} 
or calculated with the code of M.~Brack \cite{Brack}.
The interactions using \Ecmmic\ have systematically smaller surface
coefficients \asurf\ than most of the other
interactions. This hints at a correlation between the 
scheme for \cmc\ and the surface tension of an effective
interaction, which becomes obvious when looking at pairs of effective 
forces which are fitted in exactly the same way but use different 
schemes for the \cmc: the non--relativistic \Es\ and \Zs, 
SLy4 and SLy6 and the relativistic NL1 and NL--Z. In all cases the
interaction employing \Ecmmic\ has
a value of \asurf\ smaller by roughly 1.0 MeV compared to the
interaction employing a simpler scheme for the \cmc.

Although the Skyrme force SkM* employs the simple scheme (C) for 
the \cmc, it gives also a rather low value for \asurf. 
As already mentioned, SkM* is an exception among the interactions 
discussed here because it was aimed at giving better \asurf\
via fitting the fission barrier of ${}^{240}$Pu 
after it was found that all earlier Skyrme interactions give wrong 
values for the surface coefficient \cite{SkM*,Brack85}.
SkM* still serves as a reference for the surface properties of 
effective interactions for self--consistent models. Similar 
emphasis on proper INM and SINM properties was taken for SkP \cite{SkP},
again resulting in a low value for \asurf.
All Skyrme interactions employing \Ecmmic\ for the \cmc, however, give
a similar (small) value for \asurf\ as SkM* which leads to the conclusion 
that effective interactions using the microscopic \cmc\
give a reasonable value for \asurf\ \emph{without} taking explicit
information on the surface coefficient into account when fitting
them while interactions with simpler \cmc\ need additional information
on the surface tension to give proper surface properties.

This is confirmed when looking at another pair of similarly 
fitted interactions: the Gogny forces \DO\ and \DOs. Both are 
fitted with the diagonal approximation \Ecmdir\ for \cmc, scheme (C), 
but \DOs\ is a refit of the original interaction taking again explicit 
information on the deformation energy of heavy nuclei into account
\cite{Ber89a}.

In the remaining part of the paper we want to investigate the reason 
for that correlation and give some typical examples for the influence
of the difference in the \cmc\ and the surface coefficient on
observables. To that end, we have chosen from all parametrisations  
discussed above a pair of Skyrme interactions, i.e.\ SLy4 and SLy6,
and a pair of RMF forces, i.e.\ NL1 and NL--Z, which are fitted 
in the same way but use different levels of approximation for the 
\cmc.  The first pair compares scheme (C) with scheme (B) while 
the second pair compares schemes (E) and (B).
As already mentioned, the correction before variation 
used for SLy6 becomes very complicated in connection with BCS states.
Therefore we follow the suggestion of \cite{SLyx} and use the \emph{a
posteriori} microscopic correction scheme (B) instead which was
already done in \cite{CwiSH,SD}. 
%
%
\subsection{Binding energy of spherical nuclei}
\label{subsect:res:spherical}
%
%
\begin{figure}[t!]      
\epsfig{file=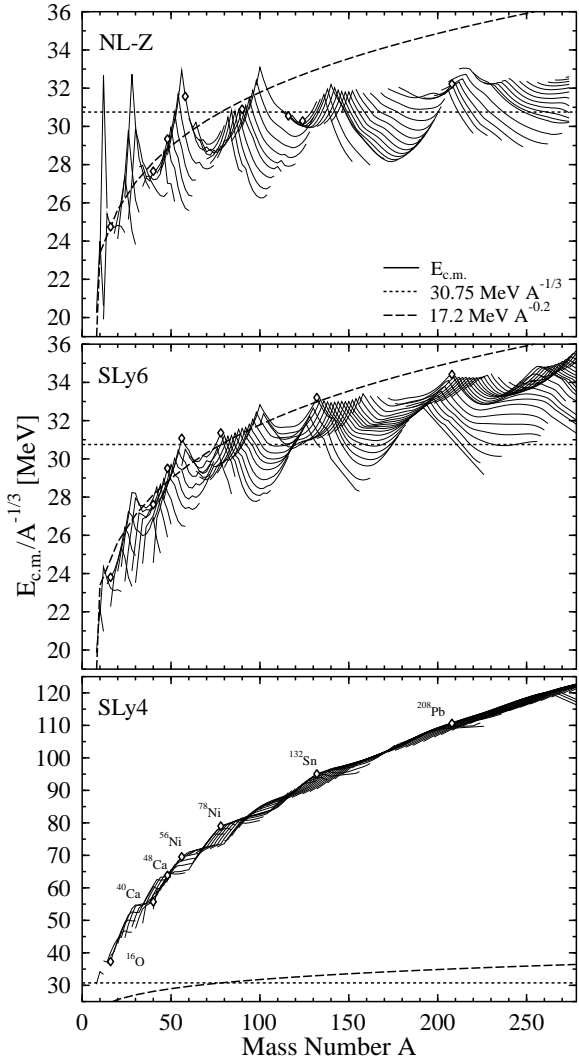}
\caption{\label{ecmsphere} 
Amplitude of the correction for center--of--mass motion \protect\Ecm\
in spherical calculations with the RMF force NL--Z (upper panel),
and the Skyrme interactions SLy6 (middle panel) and SLy4 (lower panel). 
The results are scaled with $A^{-1/3}$ to remove the smooth trend of
the microscopic \protect\cmc\ with $A$. Solid lines connect nuclei 
in isotopic chains, the dashed line is the fit (\protect\ref{eq:Ecm:fit})
used for NL1, the dotted line the harmonic oscillator estimate 
(\protect\ref{eq:Ecm:HOest}). The diamonds denote the nuclei used in 
the fit of the particular interaction.
}
\end{figure}
%
%
The fit of effective interactions for self--consistent nuclear structure
models is usually performed with spherical doubly--magic or semi--magic 
nuclei. The origin of the correlation between surface 
tension and the scheme for \cmc\ found above can be understood looking
at the systematics of the \cmc\ throughout the chart of nuclei, see 
Figure~\ref{ecmsphere}, which shows \Ecm\ drawn
versus the mass number $A$ calculated
in spherical symmetry for even--even nuclei between 
the (calculated) two--nucleon drip--lines.

Looking at the results for NL--Z, one sees that 
the \cmc\ shows pronounced shell effects and has maximum values at shell 
closures. \Ecmmic\ is a measure for the localisation of the many--body 
wave function and in magic nuclei the density distribution is indeed 
somewhat more localised: rms-radii and surface thicknesses are 
smaller than the value obtained from interpolating between adjacent nuclei, 
which in turn leads to a larger value
for the variance of the total momentum \Psqr. It is to be noted that the
results from calculations in spherical symmetry plotted in 
Fig~\ref{ecmsphere} may not be taken too seriously when one is far away
from proton and neutron shell closures: there the nuclei can be 
expected to be deformed with a more localised density distribution
and somewhat larger \Ecm. However, this does not affect the 
conclusions drawn here.

For heavy nuclei \mbox{$A > 100$} \Ecmmic\ oscillates around \Ecmosc, 
while \Ecmfit\ gives a better approximation to \Ecmmic\
for small nuclei up to ${}^{90}{\rm Zr}$, 
but overestimates \Ecmmic\ of heavy nuclei. It is noteworthy that none 
of the algebraic schemes for the \cmc\ used so far is able to approximate 
the smooth trend of \Ecmmic\ reasonably for all nuclei throughout the 
chart of nuclei. Improved fits are of course possible, but
the shell effects visible in Fig~\ref{ecmsphere} 
can be described only when \Ecmmic\ is employed for the \cmc.

At a first glance the results for SLy6 look very similar to 
those obtained with NL--Z, but there are 
some significant differences: The shell oscillations of \Ecmmic\
for small nuclei are much smaller in the SHF than in the RMF and 
\Ecmmic\ from SLy6 decreases more slowly with increasing mass 
number than \Ecmmic\ from NL--Z. This explains why \Ecmfit\ 
used for the RMF force NL1 is so far off \Ecmmic\ for NL--Z: 
The fit (\ref{eq:Ecm:fit})
was performed using values of \Ecmmic\ calculated with
the Skyrme interaction \Zs\ which gives results
very similar to those for SLy6 but which are again too large 
compared to values calculated from RMF model wave functions.

The results for \Ecmdir\ used for SLy4 are qualitatively and 
quantitatively very different compared to \Ecmmic\ 
used with NL--Z and SLy6: the shell fluctuations have nearly vanished
and -- even more important -- the diagonal part \Ecmdir\ does 
not scale with $A^{-1/3}$. This demonstrates that the exchange 
contributions to \Ecmmic\ are of the same order as the direct terms
-- but with opposite sign -- and cancel them nearly which gives 
the $A$ dependence of \Ecmmic\ found for NL--Z and SLy6. 
%
%
\begin{table}[t]   
\caption{\label{Tab:Ecm:LDM}
Coefficients  $\tilde{a}_{\rm vol}$ and $\tilde{a}_{\rm surf}$
of the liquid--drop expansion (\protect\ref{eq:DeltaEcm}) of the 
difference between \Ecmmic\ and \Ecmdir\ calculated for the fit 
nuclei of some Skyrme forces which use the diagonal
approximation scheme (C). For those parameterisations where
an equally fitted interaction with full microscopic \cmc\
exists, $\Delta \avol$ ($\Delta \asurf$) gives
the difference between the volume (surface) coefficients
of those interactions, e.g.\ \mbox{$\Delta \asurf (\text{SLy4}) = 
\asurf (\text{SLy4}) - \asurf (\text{SLy6})$}. Most of the
Skyrme interactions using scheme (C) which are listed 
in Table~\protect\ref{NucMat} are mainly fitted to nuclear matter
properties (and to a few finite nuclei only) and therefore
omitted.
}
%
%
\begin{center}
\begin{tabular}{lcc|lcc}
\hline\noalign{\smallskip}
Force  & $\tilde{a}_{\rm vol}$
       & $\tilde{a}_{\rm surf}$
       &
       & $\Delta \avol$
       & $\Delta \asurf$ \\
       & $[{\rm MeV}]$
       & $[{\rm MeV}]$
       & 
       & $[{\rm MeV}]$  
       & $[{\rm MeV}]$ \\
\noalign{\smallskip}\hline\noalign{\smallskip}
SkM*   & $-0.14$ & 1.2  & --          &  --     & --    \\
\Es    & $-0.13$ & 1.2  & \Es$-$\Zs   & $-0.15$ & 1.3   \\ 
SLy4   & $-0.15$ & 1.2  & SLy4$-$SLy6 & $-0.05$ & 0.8   \\
\noalign{\smallskip}\hline\noalign{\smallskip}
\end{tabular}
\end{center}
\end{table}
%
%

The wrong trend of \Ecmdir\ (as compared to \Ecmmic) with 
$A$ has to be compensated by the 
effective interaction ${\cal E}_{\rm int}$ in order to obtain
the proper binding energy of the (fit) nuclei.
This is the key to understand the large differences 
in the surface tension found in Sect.~\ref{subsect:res:nucmat}:
The error in the  $A$ dependence of simple schemes for \cmc\ puts
a wrong $A$ dependence into the effective interaction
which reveals itself in the nuclear matter properties.

It is instructive to fit the difference 
$\Delta \Ecm = \Ecmdir - \Ecmmic$ between the two
schemes for \cmc\ for the nuclei used in the fit of the 
interactions SkM*, \Es, and SLy4, with the simple mass formula
\begin{equation}
\label{eq:DeltaEcm}
\Delta \Ecm  
=    \tilde{a}_{\rm vol}   \, A 
   + \tilde{a}_{\rm surf}  \, A^{2/3}
\quad .
\end{equation}
The resulting values for the parameters $\tilde{a}_{\rm vol}$ and 
$\tilde{a}_{\rm surf}$ are given in Table~\ref{Tab:Ecm:LDM}
together with the actual differences $\Delta \avol$ and $\Delta \asurf$
between the nuclear matter properties of pairs the similar fitted 
interactions \Es\ and \Zs, respectively SLy4 and SLy6.
First of all it is remarkable that the coefficients $\tilde{a}_{\rm vol}$
and $ \tilde{a}_{\rm surf}$ are very similar for all interactions,
which means that the amplitude of the \cmc\ is very similar
for all non--relativistic interactions.
In the ideal case the nuclear matter properties of an effective 
interaction fitted with a simple \cmc\ would be larger by the values 
of the coefficients of (\ref{eq:DeltaEcm}) than the
nuclear matter properties of a force fitted in the same way but 
using \Ecmmic. This is indeed the case for \Es\
and \Zs\, where \mbox{$\tilde{a}_{\rm vol} \approx \Delta \avol$}
and \mbox{$\tilde{a}_{\rm surf} \approx \Delta \asurf$}. However,
for other interactions things are complicated by the fact that 
several nuclear matter properties were constrained during the fit, 
hiding (but not curing) the problems caused by the simple schemes 
for \cmc. 
The constraint on \avol\ for SLy4 and SLy6 diminishes the difference 
$\Delta \avol$ and probably with that also the difference $\Delta \asurf$ 
but one can expect that there are instead larger differences than 
necessary in the higher--order terms of the liquid--drop expansion 
since $\Delta \Ecm$ itself 
is not affected by constraints on nuclear matter properties.
Including symmetry terms in (\ref{eq:DeltaEcm})
neither improves the quality of the fit nor significantly changes 
the values of $\tilde{a}_{\rm vol}$ and $\tilde{a}_{\rm surf}$.
This simple picture works well to explain how the use 
of the approximate \cmc\ scheme (C) affects nuclear 
matter properties of non--relativistic interactions. 
The fits of $\Delta \Ecm$ with (\ref{eq:DeltaEcm}) for 
these interactions give an average deviation of
$\delta \Delta E_{\rm c.m.}^2 \approx 0.09 \, \MeV^2$.
%
%
\begin{table}[t]   
\caption{\label{Tab:Ecm:LDM2}
Coefficients of the liquid--drop expansion (\protect\ref{eq:DeltaEcm2}) 
of the difference between \Ecmmic\ and the approximation to it used 
for some relativistic forces. To obtain a fit of the same quality 
as in case of the Skyrme forces discussed above a curvature term has 
to be included.
}
\begin{center}
\begin{tabular}{lcccc}
\hline\noalign{\smallskip}
Force  & $\tilde{a}_{\rm vol}$
       & $\tilde{a}_{\rm surf}$
       & $\tilde{a}_{\rm curv}$ \\
       & $[{\rm MeV}]$
       & $[{\rm MeV}]$
       & $[{\rm MeV}]$ \\
\noalign{\smallskip}\hline\noalign{\smallskip}
NL1    & $-0.12$ & $1.2$ & $-3.2$ \\
NL3    & $-0.11$ & $1.1$ & $-2.7$ \\
NL--SH & $-0.10$ & $1.0$ & $-2.4$ \\
\noalign{\smallskip}\hline\noalign{\smallskip}
\end{tabular}
\end{center}
\end{table}
%
%

For the RMF, however, the usual approximations are schemes (D) and (E).
Here, the mass formula (\ref{eq:DeltaEcm}) gives very bad fits of 
$\Delta \Ecm$ with $\delta \Delta E_{\rm c.m.}^2 \approx 0.6 \, \MeV^2$
for all interactions. To obtain a fit of the same quality as in the 
case of non--relativistic interactions a curvature term has to be included
\begin{equation}
\label{eq:DeltaEcm2}
\Delta \Ecm  
=    \tilde{a}_{\rm vol}   \, A
   + \tilde{a}_{\rm surf}  \, A^{2/3}
   + \tilde{a}_{\rm curv}  \, A^{1/3}
\quad ,
\end{equation}
which yields again an acceptable error of 
$\delta \Delta E_{\rm c.m.}^2 \approx 0.09 \, \MeV^2$. The actual values 
of the coefficients $\tilde{a}_i$ are listed in Table~\ref{Tab:Ecm:LDM2}.
The volume and surface coefficients are of similar size as in case
of the SHF model. In the RMF we have a pair of similarly fitted
forces, namely NL1 with the recipe (E) and NL--Z with the microscopic
\cmc\ (B). And again, the differences of nuclear matter properties 
between these forces are quite close to the coefficients $\tilde{a}_i$ 
for NL1: \mbox{$\tilde{a}_{\rm vol} \approx \Delta \avol = -0.23$} MeV 
and \mbox{$\tilde{a}_{\rm surf} \approx \Delta \asurf = 1.0$} MeV.
Altogether we find that the RMF behaves much similar as the SHF 
concerning the impact of the \cmc\ on extrapolation to nuclear 
matter properties.

One may have the impression that the correlation between the recipe
for the \cmc\ and the emerging surface tension is a merely technical
problem, but one has to be aware that the choice of the recipe has
remarkable consequences for extrapolations. Besides the different 
extrapolation of total binding energies visible in Fig.~\ref{ecmsphere}, 
we discuss in what follows the dramatic consequences for strongly 
deformed systems.
%
%
\subsection{Deformation Energy of Heavy Nuclei}
We have already seen that a different treatment of the \cmc\
in the fit of an effective interaction leads to very different surface
properties even when the interactions are otherwise fitted in
exactly the same way. To get an impression of the amplitude of this
effect, we now look at superdeformed states and fission
barriers of heavy nuclei, where experimental information about the 
deformation energy is available even for very large deformations. As an
example we have chosen $^{240}{\rm Pu}$ which provides the standard 
testing ground for the capability of self-consistent nuclear mean--field 
models to describe fission barriers, see e.g.\ 
\cite{SkM*,Ber89a,Flo74a,Gir83a,Blu94a,Asymrmf}.

In the following we will present deformation energy curves calculated
in axial symmetry, allowing for reflection asymmetric shapes with a 
damped quadrupole constraint; for numerical details see \cite{Asymrmf}. 
The deformation energy curves are shown versus the dimensionless 
quadrupole moment of the mass density which is defined as
\begin{displaymath}
\beta_2
= \frac{4 \pi}{3 A r_0^2} \; \langle r^2 \, Y_{2 0} \rangle 
\quad \hbox{with $r_0 = 1.2 \, A^{1/3} \; $fm}
\end{displaymath}
and which has to be distinguished from the generating deformation parameter 
which is used in macroscopic--micro\-scopic models \cite{beta}. 
%
%
\begin{figure}[t!]      
\epsfig{file=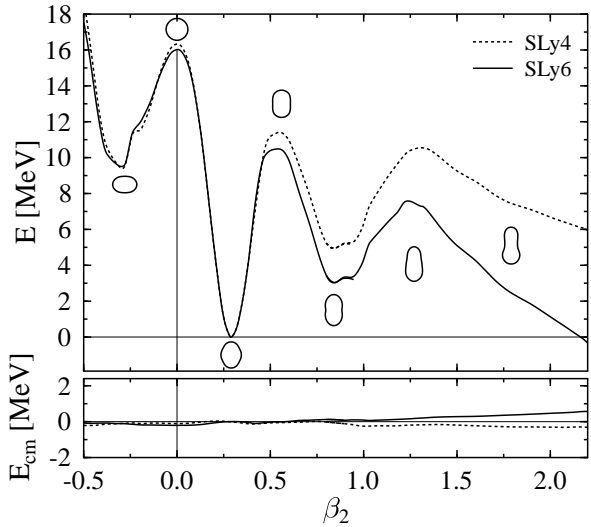}
\caption{\label{146094shf} 
  Deformation energy (normalised to the prolate ground state) 
 of ${}^{240}{\rm Pu}$ calculated with the Skyrme
  interactions SLy4 and SLy6. The lower panel shows 
  the change of the \cmc\ compared to the ground--state value
  in the same scale as the deformation
  energy in the upper panel. The various shapes along the fission 
  path are indicated by the contours of the density at 
  \mbox{$\rho_0 = 0.07 \, {\rm fm}^{-3}$}. 
}
\end{figure}
%
%

The upper panel of Fig.~\ref{146094shf} shows the deformation energy of
${}^{240}{\rm Pu}$ calculated in the SHF model 
with the interactions SLy4 and SLy6. The first barrier is slightly 
lowered for triaxial shapes which are not considered here,
but this has no influence on the conclusions drawn in this paper. The 
authors of \cite{SD}, however, report a reduction of the first barrier 
of 2.1 MeV for SLy4 when allowing for triaxial shapes. For deformations 
smaller than the value at the isomeric state at $\beta_2 = 0.85$, 
the minimum configurations turn out to be reflection symmetric, 
while for larger deformations the fission path prefers asymmetric shapes.

Both Skyrme interactions give very similar deformation energy at small 
deformations inside the first barrier, but beyond the first barrier
a significant difference between the deformation energy curves of SLy4 
and SLy6 becomes visible which increases steadily with deformation. At the 
second, superdeformed (SD) minimum the difference is already 
1.9 MeV and increases to 6.3 MeV at \mbox{$\beta_2 = 2.2$}. 
SLy4 gives a much broader and higher fission barrier than SLy6 which 
will make a huge difference when calculating fission half-lives.
The difference between the interactions is not caused by the variation 
of the \cmc\ with deformation as can be seen in the lower panel of
Fig.~\ref{146094shf}; it is caused by the difference in \asurf\ between 
the two interactions. This was already noticed by the authors
of Ref.\ \cite{SLyx}.

The excitation energy of the SD minimum is determined mainly 
by the interplay two very different effects: The variation 
of surface (and surface 
symmetry) energy with deformation contained in the bulk properties 
of the effective interaction (the Coulomb energy which 
varies with deformation as well is nearly the same for all
interactions) and the variation of shell effects with deformation
(which is fixed by the shell structure at spherical shape). 
SLy4 and SLy6 give nearly the same single--particle energies 
for spherical nuclei, 
the fission paths are identical, the difference visible in 
Fig.~\ref{146094shf} is caused only by the difference in
surface tension between the two interactions.

This gives a possible explanation for the finding of the authors of 
\cite{SD} that the Skyrme interaction SLy4  
overestimates the excitation energy of the SD minima in nuclei around 
\mbox{$A \approx 190$} and \mbox{$A \approx 240$} while it reproduces
nicely the separation energies within the first and SD minima. In each
region either all ground--states and all SD minima have roughly the 
same deformation, therefore the separation energy calculated between
states in the same well is not affected by the surface tension.
This might explain also to some extent the preference for SkM*, 
SkI3 and SkP for the description of superdeformed states in 
${}^{194}{\rm Hg}$ found in \cite{Tak98a}. 
All three forces have rather small surface coefficients \asurf\
(see Table~\ref{NucMat} and Fig.~\ref{asurf}),
SkI3 because it employs the microscopic \cmc, SkM* and SkP
because the value of \asurf\ was constrained during the fit 
of these particular interactions.
However, the excitation energy of superdeformed states
is determined by the interplay of surface properties and
shell structure, both have to be properly described to
reproduce experimental data throughout the whole chart of nuclei.
More investigation in that direction is needed.
%
%
\begin{figure}[t!]      
\epsfig{file=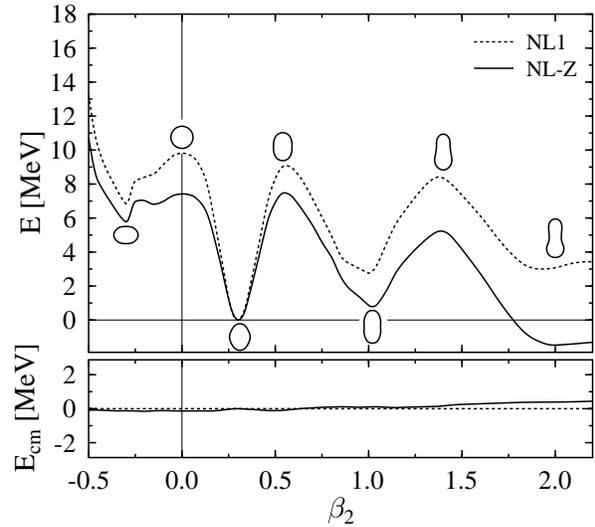}
\caption{\label{146094rmf} 
  Fission barrier of ${}^{240}{\rm Pu}$ calculated with the RMF forces 
  NL1 and NL--Z drawn in the same manner and using the same scales 
  as in Fig.~\protect\ref{146094shf}.
}
\end{figure}
%
%

Figure~\ref{146094rmf} shows the same as Fig.~\ref{146094shf},
but for the RMF forces NL1 and NL--Z. The qualitative features 
of the fission barrier are the same as those found for the Skyrme
forces, but there are significant quantitative differences. 
Comparing both figures it can be clearly seen that all barriers are 
much smaller for the RMF forces compared to the Skyrme interactions. 
Again, the RMF force which employs the ``cheaper'' \cmc\ gives 
the larger deformation energy compared to the ground-state value. 
There are now some differences between NL1 and NL--Z at small deformations
as well, but for large prolate deformations we regain our finding for
the Skyrme forces: The difference in the deformation energy increases with
deformation, it is already $1.9$ MeV at the SD minimum and $4.7$ MeV
around \mbox{$\beta_2 = 2.2$}.
The RMF forces shift the SD minimum to a slightly larger
deformation $\beta_2 \approx 1.0$ than the SLy$x$ forces. For NL--Z
the excitation energy of the SD minimum of $-0.8$ MeV is 
definitively too small compared to the experimental value of 2.4 MeV,
while the value of 2.8 MeV predicted by NL1 is quite close. 
We expect, however, that the better description of the SD state 
obtained with NL1 is accidental because the surface coefficient 
of this interaction is definitively too large.

Again the difference of the deformation energy curves is not caused
by the actual contribution of the \cmc\ to the binding energy,
see the lower panel of Fig.~\ref{146094rmf}. \Ecmmic\ 
varies only by a few 100 keV along the fission path
while the value for \Ecmfit\ is of course constant.

We have to revise our conclusions from \cite{Asymrmf} where the fission
barriers of ${}^{240}{\rm Pu}$ calculated with PL--40 \cite{PL40}
and NL1 were compared. Both interactions are fitted in the same way
but differ in two details: the treatment of the \cmc\  -- where PL--40 
employs the same scheme as NL--Z  -- and the density dependence, where
NL1 and NL--Z employ the standard non--linear self-interaction of the
scalar field while PL--40 uses a stabilised self-interaction, see 
\cite{Rei89a,PL40} for details. We have concluded in 
Ref.~\cite{Asymrmf} that the difference 
between the fission barriers is caused by the difference in the density
dependence of the two forces PL--40 and NL1, but as can be seen
comparing Fig~\ref{146094rmf} with the results of \cite{Asymrmf}
NL--Z and PL--40 give very similar fission barriers, 
therefore the difference between NL1 and PL--40 is caused by 
the different treatment of the \cmc\ during the fit of these 
interactions. This means in turn that using the stabilised 
non--linearity instead of the standard non--linearity in the RMF 
has only a very small influence on the fission barriers of
heavy nuclei. 

It is to be noted that the mere \cmc\ (\ref{eq:Ecm:mic}) becomes
inconsistent in the asymptotics of fission (i.e.\ beyond the 
scission point). One still would use one common \Ecm\ whereas
the two fragments should now acquire each their own \Ecm. The 
problem can be resolved by including also the correction for
spurious quadrupole vibrations \cite{GCMrev,Urbano}. But this
complication can be neglected when looking at the fission 
barrier of heavy nuclei where the scission point is far beyond 
the fission point.
%
%
\subsection{Deformation Energy of Light Nuclei}
%
%
\begin{figure}[t!]      
\epsfig{file=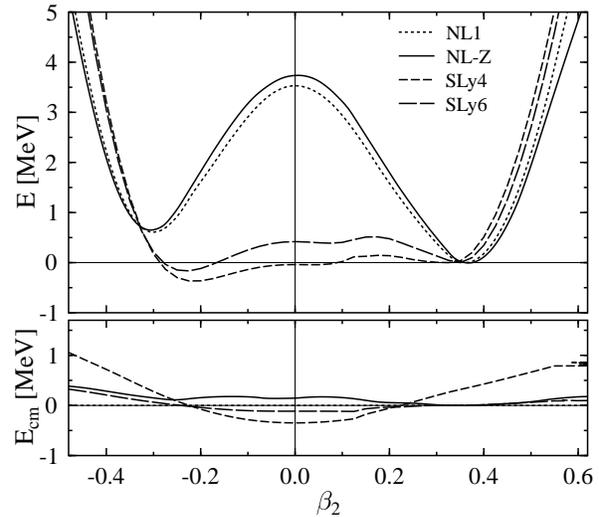}
\caption{\label{028016} 
  Deformation energy (upper panel) and variation of the \cmc\
  (lower panel) of ${}^{44}{\rm S}$ calculated with NL1, NL--Z,
  SLy4 and SLy6. The deformation energy $E$ and the \cmc\ \Ecm\ are 
  normalised to their values at the prolate minimum.
  The two panels use again the same energy scale.
}
\end{figure}
%
%
Figure~\ref{028016} shows the deformation energy for the nucleus 
${}^{44}{\rm S}$ calculated again with SLy4, SLy6, NL1, and NL--Z. 
The potential landscapes for the relativistic and non--relativistic 
interactions look very different: SLy4 and SLy6 predict a rather 
soft potential energy surface with a shallow oblate minimum, while
NL1 and NL--Z predict an energy surface with deep prolate and oblate
minima of comparable binding energy (which might be connected
by triaxial configurations). The main difference between the 
models appears at small deformations around \mbox{$\beta_2 = 0$}, while
all interactions give similar predictions for the deformation of the 
prolate and oblate minima respectively. This is interesting for itself
as it has dramatic influence on the excitation spectrum of
this nucleus, but we do not want to discuss this further here
and refer to \cite{ShapeCoex} and references therein.

In contrast to the heavy system ${}^{240}{\rm Pu}$ there now appear 
visible differences at small deformations as well, which are directly related 
to the different variation of \Ecmdir\ and \Ecmmic\ with deformation, see 
the lower panel of Fig.~\ref{028016}. Comparing SLy4 and SLy6, one sees
that the diagonal part of the \cmc\ oscillates 
with larger amplitude than the full correction including exchange terms.
The relativistic forces NL1 and NL--Z show a small difference at 
small deformations as well, but this difference has another origin: 
\Ecmmic\ is slowly varying while the fit value 
is simply constant. At large prolate deformations the difference 
in surface tension between SLy4 and SLy6 gives again a large difference 
between the potential energy surfaces.
The differences caused by the \cmc\ among these particular 
relativistic and non--relativistic forces are rather small compared to the 
difference between the predictions of the non--relativistic and 
relativistic models and it is to be noted that SLy4 and SLy6 are
by no means representative for the predictions of Skyrme
interactions, see \cite{ShapeCoex}. 
However, the differences we see for small nuclei at small deformations
caused by differences in the variation of the \cmc\ with deformation 
are subtle compared to the huge effect caused by the 
difference in surface tension seen at large deformations.
%
%
\section{Summary}
\label{sec:conclusion}
We have discussed the various approximate schemes for performing the
\cmc\ in connection with self--consistent mean--field models. In
particular, we have scrutinised the effect of the actual recipe for
the \cmc\ used while fitting a parametrisation. The basic point is
that the different recipes differ in their trends with mass number $A$
and isospin. Mismatches thus inherent in a recipe are counterweighted
to a certain extend by (automatic) readjustments of the model
parameters. This, in turn, builds into the forces different properties
concerning extrapolations.

We have considered the \cmc\ of order \Psqr\ throughout. The full evaluation 
of \Ecmmic\ was compared with the approximation through the diagonal part of 
that expression and with a simple estimate $\propto A^{-\gamma}$.
The comparison of the \cmc\ as such shows significant differences
between these three schemes. The full correction shows considerable
shell effects which are completely absent in all simple estimates. 
Moreover, the average trends of \Ecm\ with $A$ and $I$ of these three 
recipes differ, which has consequences for the 
properties of effective interactions fitted with simple schemes for \cmc.
We see a strong effect on the surface
energy where the forces with approximate \cmc\ (diagonal part or
simple estimate) give systematically larger values than
interactions employing \Ecmmic.

This error in surface energy leads to a much different evolution of
energies with deformation. As a consequence, fission barriers in actinide
nuclei are about 4 MeV larger for those forces which employed the
approximate recipes during their fit. Furthermore, predictions of the
relative height of deformation isomers and collective vibrations
therein will come out significantly different. For light nuclei -- where
the relative importance of the \cmc\ is larger -- one even has to be
aware of differences at small deformations. Altogether, we strongly
recommend to use consistently the full correction
$\Psqr / (2mA)$ and forces derived with that
recipe. Any further approximation beyond that (already approximate)
stage induces systematic errors in extrapolation to heavy nuclei,
large deformation, and probably also to nuclei far off the stability
line.
%
%
\section*{Acknowledgements}
The authors thank W. Nazarewicz for useful comments.
This work was supported by Bundesministerium f\"ur Bildung 
und Forschung (BMBF), Project No.\ 06 ER 808 and by 
the U.S.\ Department of Energy under Contract No.\ DE--FG02--96ER40963 
with the University of Tennessee and Contract No.\ DE--FG02--97ER41019 
with the University of North Carolina and by the NATO grant SA.5--2--05 
(CRG.971541). The Joint Institute for Heavy Ion Research has as member 
institutions the University of Tennessee, Vanderbilt University, 
and the Oak Ridge National Laboratory; it is supported by the members 
and by the Department of Energy through contract No.\ DE--FG05--87ER40361 
with the University of Tennessee.
%
%
\begin{appendix}
\section{The Calculation of $\langle\hat{\bf P}_{\rm c.m.}^2\rangle$}
In this appendix we present the formulae needed to calculate the
expectation value of \Psqr\ in relativistic and non--relativistic
models assuming spherical, axial symmetry or triaxial symmetry 
for a BCS many--body state. Expressing (\ref{eq:BCSPsqr}) in 
terms of matrix elements $\Delta_{\alpha \alpha}$
of the Laplacian and  $\nabla_{\alpha \beta}$ of the nabla operator
respectively, one obtains
\begin{eqnarray}
\label{eq:Ecm:mic2}
\lefteqn{
\Psqr
} \nn \\
& = & - \hbar^2
        \Big[ \sum_{\alpha \gk 0} v_\alpha^2 \; \Delta_{\alpha \alpha} 
      + \! \sum_{\alpha,\beta \gk 0} \!
        \big(    v_\alpha^2 \, v_\beta^2 \,
                 \nabla_{\alpha \beta} \cdot \nabla^{*}_{\alpha \beta}
      \nonumber \\
&   & \quad
              -  v_\alpha  u_\alpha  v_\beta u_\beta \,
                 \nabla_{\alpha \beta} \cdot 
                 \nabla_{\bar\alpha \bar\beta} 
        \big)
        \Big]
\quad .
\end{eqnarray}
In case of a time--even many--body state this simplifies with
$\Delta_{\bar\alpha \bar\alpha} = \Delta_{\alpha \alpha}$,
$\nabla_{\bar\alpha \bar\beta} = \nabla^*_{\alpha \beta}$ and
$\nabla_{\bar\alpha \beta} = - \nabla^*_{\alpha \bar\beta}$ to
\begin{eqnarray}
\label{eq:Ecm:mic2b}
\Psqr
& = & - 2 \hbar^2
        \Big[ \sum_{\alpha > 0} v_\alpha^2 \; \Delta_{\alpha \alpha} 
      + \! \sum_{\alpha,\beta>0} \! 
        v_\alpha v_\beta ( v_\alpha v_\beta + u_\alpha u_\beta ) 
      \nonumber \\
&   & \quad \times
        \Big(  | \nabla_{\alpha \beta} |^2
             + | \nabla_{\alpha \bar{\beta}} |^2 \Big)
        \Big]
\quad .
\end{eqnarray}
%
%
\subsection{Non--Relativistic Models}
\subsubsection{Spherical Symmetry}
In spherical symmetry non--relativistic spinors are given by
\begin{equation}
\Psi_{\alpha j \ell m} (\vec{r})
= \langle \vec{r} | \alpha j \ell m \rangle
= \psi_{\alpha j \ell} (r) \, \Omega_{j \ell m} (\vartheta, \phi)
\end{equation}
where $\alpha$ is the principal quantum number, $j$ the total angular
momentum, $\ell$ the orbital angular momentum, and $m$ the projection
of the total angular momentum. $\psi_{\alpha j \ell}$ is the radial
wave function to be treated numerically, while the spinor spherical
harmonics $\Omega_{j \ell m}$ \cite{Var88a} can be treated analytically.
When calculating the \cmc\ (\ref{eq:Ecm:mic2b}) 
\begin{eqnarray}
\label{eq:Ecm:mic:sphere}
\lefteqn{
\langle \vec{P}^2_{\rm c.m.} \rangle
} \nn \\
& = & - \hbar^2 \sum_{\alpha, j, \ell} 
        v_{\alpha j \ell}^2 \; 
        \sum_{m = -j}^{+j} 
        \langle \alpha j \ell m | \Delta | \alpha j \ell m \rangle 
      \nn \\
&   & - \hbar^2 \! \! \! \sum_{\alpha, \beta, j, J , \ell, L} \! \! \! 
        v_{\alpha j \ell}  \, v_{\beta J L} \,
       (   v_{\alpha j \ell} \, v_{\beta J L} 
         + u_{\alpha j \ell} \, u_{\beta J L} ) 
      \nn \\
&   & \quad \times
      \sum_{m = -j}^{+j} \sum_{M = -J}^{+J}
      \big| \langle \alpha j \ell m | \nabla  | \beta J L M \rangle \big|^2
\end{eqnarray}
the summation over $m$ and $M$ can be performed analytically: the sum 
over $m$ (for given $j$ and $\ell$) of the matrix elements of the 
Laplacian gives simply $2j+1$ times the expectation value of 
the Laplacian of $\psi_{\alpha j \ell}$, while the square of the 
matrix elements of the nabla operator reads
\begin{eqnarray}
\lefteqn{
\label{eq:sphere:nablame}
\sum_{m = -j}^{+j}
\sum_{M = -J}^{+J}
\big| \langle \alpha j \ell m | \nabla  | \beta J L M \rangle \big|^2
} \nonumber \\
& = & (-1)^{L+\ell} \, (2J+1) \, (2j+1) \,
     \left\{ \begin{array}{ccc} J    & j & 1 \\
                                \ell & L & \tfrac{1}{2}
             \end{array}
     \right\}^2
     \nonumber \\
&   & \times
      \big[
          L \, \delta_{\ell, L-1} \, 
          A_{\alpha \beta} \, B_{\beta \alpha}
        + \ell \, \delta_{L, \ell-1} \, 
          A_{\beta \alpha}  \, B_{\alpha \beta}
      \big]
\end{eqnarray}
with
\begin{eqnarray}
A_{\alpha \beta}
& = & \int \! \rmd r \; r^2 \;
      \psi_{\alpha j \ell} \,
      \Big( \partial_r + \tfrac{L+1}{r} \Big) \,
      \psi_{\beta J L}
      \quad ,
      \nonumber \\
B_{\alpha \beta}
& = & \int \! \rmd r \; r^2 \;
      \psi_{\alpha j \ell} \,
      \Big( \partial_r - \tfrac{L}{r} \Big) \,
      \psi_{\beta J L}
\quad .
\end{eqnarray}
The square of the $6j$ symbol occurring in (\ref{eq:sphere:nablame}) 
is given by
\begin{eqnarray}
\label{eq:6j}
\left\{ \begin{array}{ccc} J    & j & 1 \\
                           \ell & L & \tfrac{1}{2}
        \end{array} \right\}^2
& = &   \frac{(\ell+J+\frac{5}{2})(\ell+J-\frac{1}{2})}
             {(2j+1)(2J+1)(2\ell+1)(2L+1)} \,
        \nn \\
&   & \quad \times
        (   \delta_{j, \ell + 1} \, \delta_{J, L + 1}
          + \delta_{j, \ell - 1} \, \delta_{J, L - 1} ) 
        \nn \\
&   & + \frac{(J-\ell+\frac{3}{2})(\ell-J+\frac{3}{2})}
             {(2j+1)(2J+1)(2\ell+1)(2L+1)} \,
         \nn \\
&   & \quad \times
       (   \delta_{j, \ell + 1} \, \delta_{J, L - 1}
          + \delta_{j, \ell - 1} \, \delta_{J, L + 1} ) 
\end{eqnarray}
%
%
\subsubsection{Axial Symmetry}
Assuming axial symmetry, a spinor with angular momentum projection $m$
is given by
\begin{equation}
\Psi_\alpha (\vec{r})
= \left( \begin{array}{c}
         \psi_\alpha^{(+)} (r,z) \,
         \exp \big[ \iunit \big( m - \tfrac{1}{2} \big) \phi \big] \\
         \psi_\alpha^{(-)} (r,z) \,
         \exp \big[ \iunit \big( m + \tfrac{1}{2} \big) \phi \big]
         \end{array}
  \right)
\quad .
\end{equation}
The $\psi_\alpha^{(\sigma)}$ can be chosen to be real.
The matrix elements of the nabla operator needed to calculate 
(\ref{eq:Ecm:mic2}) or (\ref{eq:Ecm:mic2b}) are given by
\begin{eqnarray}
\label{eq:ax:nablame}
\nabla_{\alpha \beta} \cdot \nabla^{*}_{\alpha \beta}
& = & \delta_{m_\alpha m_\beta} \, A_{\alpha \beta} \, A_{\alpha \beta}
      \nn \\
&   & 
      + \tfrac{1}{2}
        \Big(  \delta_{m_\alpha, m_\beta-1} \, B_{\alpha \beta}^- \, B_{\alpha \beta}^-
             + \delta_{m_\alpha, m_\beta+1} \, B_{\alpha \beta}^+ \, B_{\alpha \beta}^+
        \Big)
      \nn \\
\nabla_{\alpha \beta} \cdot \nabla_{\bar\alpha \bar\beta} 
& = & \delta_{m_\alpha m_\beta} \, A_{\alpha \beta} \, A_{\bar\alpha \bar\beta}
      \nn \\
&   & 
      + \tfrac{1}{2}
        \Big(  \delta_{m_\alpha, m_\beta-1} \, 
               B_{\alpha \beta}^- \, B_{\bar\alpha \bar\beta}^+
            + \delta_{m_\alpha, m_\beta+1} \, 
               B_{\alpha \beta}^+ \, B_{\bar\alpha \bar\beta}^-
        \Big)
      \nn \\
\nabla_{\alpha \bar{\beta}} \cdot \nabla^{*}_{\alpha \bar{\beta}} 
& = & \tfrac{1}{2} \,
      \delta_{m_\alpha, \frac{1}{2}} \,
      \delta_{m_\beta , \frac{1}{2}} \,
      C_{\alpha \beta} \, C_{\alpha \beta}
\end{eqnarray}
with
\begin{eqnarray}
A_{\alpha \beta}
& = & \int \! \rmd^3 r \sum_{\sigma = \pm}
      \psi_\alpha^{(\sigma)} \, \partial_z \, \psi_\beta^{(\sigma)}
      \\
B_{\alpha \beta}^\pm
& = & \int \! \rmd^3 r \sum_{\sigma = \pm}
       \psi_\alpha^{(\sigma)} 
       \bigg( \partial_r \pm \frac{m_\beta - \frac{\sigma}{2}}{r} \bigg)
       \psi_\beta^{(\sigma)}
      \\
C_{\alpha \beta}
& = & \int \! \rmd^3 r \sum_{\sigma = \pm} (-\sigma)
      \psi_\alpha^{(\sigma)} 
      \bigg( \partial_r + \frac{m_\beta + \frac{\sigma}{2}}{r} \bigg)
       \psi_\beta^{(-\sigma)}
\end{eqnarray}
where the volume element is given by 
$\rmd^3 r = 2 \pi \, \rmd r \, r \, \rmd z$.
%
%
\subsubsection{Cartesian Representation}
In Cartesian representation the non--relativistic spinors 
are given by
\begin{equation}
\Psi_\alpha (\vec{r})
= \left( \begin{array}{c}
           \psi_\alpha^{(++)} (\vec{r}) 
         + \iunit \, \psi_\alpha^{(+-)} (\vec{r}) \\
           \psi_\alpha^{(-+)} (\vec{r}) 
         + \iunit \, \psi_\alpha^{(--)} (\vec{r})
         \end{array}
  \right)
\quad ,  
\end{equation}
where the $ \psi_\alpha^{(\sigma \eta)}$ are real functions.
The matrix elements of the nabla operator needed for the calculation
of (\ref{eq:Ecm:mic2}) are given by
\begin{eqnarray}
\label{eq:kart:nablame}
\nabla_{\alpha \beta} \cdot \nabla_{\alpha \beta}^*
& = & \vec{A}_{\alpha \beta}^2 + \vec{B}_{\alpha \beta}^2 
      \nn \\
\nabla_{\alpha \bar\beta} \cdot \nabla_{\alpha \bar\beta}^*
& = & \vec{A}_{\alpha \bar\beta}^2 + \vec{B}_{\alpha \bar\beta}^2 
\end{eqnarray}
with
\begin{eqnarray}
A_{\alpha \beta}
& = & \int \! \rmd^3 r \! 
      \sum_{\sigma, \eta}
      \psi_\alpha^{(\sigma \eta)} \,
      \nabla \, 
      \psi_\beta^{(\sigma \eta)}
      \nn \\
A_{\alpha \bar\beta}
& = & \int \! \rmd^3 r \!
      \sum_{\sigma, \eta}
      \sigma \eta \, \psi_\alpha^{(-\sigma \eta)} \, 
      \nabla \,
      \psi_\beta^{(\sigma \eta)}
      \nn \\
B_{\alpha \beta}
& = & \int \! \rmd^3 r \! 
      \sum_{\sigma, \eta}
      \eta \, \psi_\alpha^{(\sigma -\eta)} \, 
      \nabla \,
      \psi_\beta^{(\sigma \eta)} 
      \nn \\
B_{\alpha \bar\beta}
& = & \int \! \rmd^3 r \!
      \sum_{\sigma, \eta}
      \sigma \, \psi_\alpha^{(-\sigma -\eta)} \, 
      \nabla \,
      \psi_\beta^{(\sigma \eta)}
\end{eqnarray}
%
%
\subsection{Relativistic Models}
Relativistic kinematics plays no role for nuclear ground states,
the \cmc\ can be calculated in non--relativistic
approximation in relativistic models. The main difference between 
relativistic and non--relativistic models is that the relativistic 
spinors have four components which changes the calculation of \Psqr.
%
%
\subsubsection{Spherical Symmetry}
The relativistic spinors in spherical symmetry are given by
\begin{equation}
\Phi_{\alpha \kappa m} (\vec{r})
  =  \langle \vec{r} | \alpha \kappa m \rangle
  =   \left( \begin{array}{c}
             \psi^{(+)}_{\alpha \kappa} (r) \, 
             \Omega_{\kappa, m} (\vartheta, \varphi) \\
             \iunit \psi^{(-)}_{\alpha \kappa} (r) \,
             \Omega_{-\kappa, m} (\vartheta, \varphi)
             \end{array}
      \right)
\end{equation}
The quantum number $\kappa$ is related to the total and orbital
angular momentum by $j = | \kappa| - \tfrac{1}{2}$ and
$\ell = j + \tfrac{1}{2} \tfrac{\kappa}{|\kappa|}$.
The real functions $\psi^{(\eta)}$ are the radial part
of the upper and lower components of the wave functions, the 
$\Omega_{\kappa,m}$ are spinor spherical harmonics \cite{Var88a}
with angular momenta corresponding to $\kappa$.
The variance of the total momentum is now given by 
\begin{eqnarray}
\label{eq:Ecm:mic:sphere:rmf}
\langle \vec{P}^2_{\rm c.m.} \rangle
& = & - \hbar^2 
        \sum_{\alpha, \kappa} v_{\alpha \kappa}^2 \; 
        \sum_{m = -j}^{+j}
        \langle \alpha \kappa m | \Delta | \alpha \kappa m \rangle
      \nn \\
&   & - \hbar^2 \sum_{\alpha, \beta, \kappa \Kappa}
        v_{\alpha \kappa}  \, v_{\beta \Kappa} \,
       (   v_{\alpha \kappa} \, v_{\beta \Kappa} 
         + u_{\alpha \kappa} \, u_{\beta \Kappa} ) 
      \nn \\
&   & \quad \times
      \sum_{m = -j}^{+j} \sum_{M = -J}^{+J}
      \big| \langle \alpha \kappa m | \nabla | \beta \Kappa M \rangle \big|^2
\end{eqnarray}
with
\begin{equation}
\label{eq:laplacemat:sphere:rmf}
\sum_{m = -j}^{+j}
\langle \alpha \kappa m | \Delta | \alpha \kappa m \rangle
  =   (2j + 1) 
      \sum_{\eta = \pm}
      \int \! \rmd r \, r^2 \,
      \psi^{(\eta)}_{\alpha \kappa} \,
      \Delta \,
      \psi^{(\eta)}_{\alpha \kappa} 
\end{equation}
and
\begin{eqnarray}
\label{eq:nablamat:sphere:rmf}
\lefteqn{
\sum_{m = -j}^{+j}
\sum_{M = -J}^{+J}
\big| \langle \alpha j \kappa | \nabla | \beta \Kappa M \rangle \big|^2
}     \\
& = & (2J+1) \, (2j+1) \,
      \sum_{\eta = \pm} (-1)^{L^{(\eta)}+\ell^{(\eta)}} 
     \left\{ \begin{array}{ccc} J    & j & 1 \\
                                \ell^{(\eta)} & L^{(\eta)} & \tfrac{1}{2}
             \end{array}
     \right\}^2
     \nn \\
&   & \times
      \Big[   L \, \delta_{\ell^{(\eta)}, L^{(\eta)}-1} \, 
          A_{\alpha \beta}^{(\eta)} \, B_{\beta \alpha}^{(\eta)}
        + \ell \, \delta_{L^{(\eta)}, \ell^{(\eta)}-1} \, 
          A_{\beta \alpha}^{(\eta)} \, B_{\alpha \beta}^{(\eta)}
      \Big]
      \nn
\end{eqnarray}
where $A$ and $B$ are given by
\begin{eqnarray}
A_{\alpha \beta}^{(\eta)}
& = & \int \! \rmd r \; r^2 \;
      \psi_{\alpha \kappa} \,
      \Big( \partial_r + \tfrac{L^{(\eta)}+1}{r} \Big) \,
      \psi_{\beta \Kappa}
      \quad ,
      \nonumber \\
B_{\alpha \beta}^{(\eta)}
& = & \int \! \rmd r \; r^2 \;
      \psi_{\alpha \kappa} \,
      \Big( \partial_r - \tfrac{L^{(\eta)}}{r} \Big) \,
      \psi_{\beta \Kappa}
\quad .
\end{eqnarray}
The $6j$ symbol appearing in (\ref{eq:nablamat:sphere:rmf}) is
again given by (\ref{eq:6j}).
%
%
\subsubsection{Axial Symmetry}
The relativistic spinors in axial symmetry are given by
\begin{equation}
\Phi_\alpha (\vec{r})
= \left( \begin{array}{r}
         \psi^{(++)}_\alpha (r,z) \, 
         \exp \big[ \iunit \big( m - \tfrac{1}{2} \big) \phi \big] \\
         \psi^{(+-)}_\alpha (r,z) \, 
         \exp \big[ \iunit \big( m + \tfrac{1}{2} \big) \phi \big] \\
         \iunit \, \psi^{(-+)}_\alpha (r,z) \, 
         \exp \big[ \iunit \big( m  -\tfrac{1}{2} \big) \phi \big] \\
         \iunit \, \psi^{(--)}_\alpha (r,z) \, 
         \exp \big[ \iunit \big( m + \tfrac{1}{2} \big) \phi \big]
         \end{array}
  \right)
\end{equation}
where $(r,z,\phi)$ are cylindrical coordinates and $m$ is the projection 
of the total angular momentum. The $\psi^{(\eta \sigma)}$ are real functions,
where $\eta$ denotes upper and lower components and $\sigma/2$ the 
spin projection. The matrix elements entering (\ref{eq:ax:nablame}) 
are now given by
\begin{eqnarray}
\Delta_{\alpha \alpha}
& = & \int \! \rmd^3 r \!
      \sum_{\eta, \sigma}
      \psi_\alpha^{(\eta \sigma)} \!
      \bigg[ \partial_r^2 
            + \tfrac{1}{r} \, \partial_r
            - \frac{(m-\frac{\sigma}{2})^2}{r^2}
            + \partial_z^2
      \bigg]
      \psi_\alpha^{(\eta \sigma)}
      \nn \\
A_{\alpha \beta}
& = & \int \! \rmd^3 r 
      \sum_{\eta, \sigma}
      \psi_\alpha^{(\eta \sigma)} \,
      \partial_z \,
      \psi_\beta^{(\eta \sigma)} 
      \nn \\
B_{\alpha \beta}^{\pm}
& = &  \int \! \rmd^3 r 
      \sum_{\eta \sigma}
      \psi_\alpha^{(\eta \sigma)}
      \bigg(     \partial_r
             \pm \frac{m_\beta-\frac{\sigma}{2}}{r}
      \bigg)
      \psi_\beta^{(\eta \sigma)}   
      \nn \\ 
C_{\alpha \beta}
& = & \int \! \rmd^3 r \!
      \sum_{\eta \sigma}
      \eta \, \sigma \,
      \psi_\alpha^{(\eta \sigma)}
      \bigg(   \partial_r
             + \frac{m_\beta + \frac{\sigma}{2}}{r}
      \bigg)
      \psi_\beta^{(\eta, -\sigma)}
\end{eqnarray}
%
%
\subsubsection{Cartesian Representation}
The relativistic spinors in Cartesian representation are given by
\begin{equation}
\Phi_\alpha (\vec{r})
= \left( \begin{array}{c}
       \psi^{(+++)}_\alpha (\vec{r}) + \iunit \, \psi^{(++-)}_\alpha (\vec{r}) \\
       \psi^{(+-+)}_\alpha (\vec{r}) + \iunit \, \psi^{(+--)}_\alpha (\vec{r}) \\
       \psi^{(-++)}_\alpha (\vec{r}) + \iunit \, \psi^{(-+-)}_\alpha (\vec{r}) \\
       \psi^{(--+)}_\alpha (\vec{r}) + \iunit \, \psi^{(---)}_\alpha (\vec{r})
       \end{array}
  \right)
\end{equation}
where the $\psi^{(\eta \sigma \varrho)}_\alpha$ are real functions. $\eta$
and $\sigma$ have the same meaning as in the axial case, $\varrho$ denotes
real and imaginary part of a spinor component.
The matrix elements of the Laplacian are given by
\begin{equation}
\Delta_{\alpha \alpha}
  =   \int \! \rmd^3 r \! 
      \sum_{\sigma, \eta, \varrho} 
      \psi_\alpha^{(\eta \sigma \varrho)}
      \Delta \,
      \psi_\alpha^{(\eta \sigma \varrho)}
\end{equation}
while the integrals needed to calculate the matrix elements of
the nabla (\ref{eq:kart:nablame}) are given by
\begin{eqnarray}
\vec{A}_{\alpha \beta}
& = & \int \! \rmd^3 r \! 
      \sum_{\sigma, \eta, \varrho} 
      \psi_\alpha^{(\eta \sigma \varrho)} \,
      \nabla \,
      \psi_\beta^{(\eta \sigma \varrho)}
      \nn \\
\vec{A}_{\alpha \bar\beta}
& = & \int \! \rmd^3 r \! 
      \sum_{\sigma, \eta, \varrho} 
      \psi_\alpha^{(\eta, -\sigma, \varrho)} \,
      \nabla \,
      \psi_\beta^{(\eta \sigma \varrho)}
      \nn \\
\vec{B}_{\alpha \beta}
& = & \int \! \rmd^3 r \! 
      \sum_{\sigma, \eta, \varrho} 
      \psi_\alpha^{(\eta, \sigma, -\varrho)} \,
      \nabla \,
      \psi_\beta^{(\eta \sigma \varrho)}
      \nn \\
\vec{B}_{\alpha \bar\beta}
& = & \int \! \rmd^3 r \! 
      \sum_{\sigma, \eta, \varrho} 
      \psi_\alpha^{(\eta, -\sigma, -\varrho)} \,
      \nabla \,
      \psi_\beta^{(\eta \sigma \varrho)}
\end{eqnarray}
\end{appendix}
%
%

%
%
\end{document}